\documentstyle[twoside,aps]{revtex}
\begin{document}
\draft
\title{Phase Diagram of a Two-Dimensional Neutral Classical \\ Coulomb Gas
              from a Non-perturbative sine-Gordon Expansion}
\author{Wen-Fa Lu$^{a,b,c}$, and Chul Koo Kim$^{a,c}$}
\address{$^a$ Institute of Physics and Applied Physics, Yonsei University,
Seoul 120-749, Korea \\
 $^b$ Department of Applied Physics, Shanghai Jiao Tong University,
Shanghai 200030, \ \ \\ the People's Republic of China
   \thanks{Permanent address, E-mail: wenfalu@online.sh.cn}
    \ \ \ \ \ \ \ \ \ \ \ \ \ \ \ \ \ \ \ \ \ \ \ \ \ \ \ \ \
     \ \ \ \ \ \ \ \ \ \ \ \ \ \ \ \ \ \ \ \ \ \ \ \ \ \ \ \ \ \ \ \ \\
 $^c$ Center for Strongly Correlated Materials Research,
Seoul National University, \ \ \ \ \ \ \ \ \ \\ Seoul 151-742, Korea
    }
\maketitle

\begin{abstract}
Within the sine-Gordon formalism of a two-dimensional neutral classical
Coulomb gas, a convergent expansion with non-perturbative nature is performed
to calculate the thermodynamic potential and construct the phase diagram. It
is shown that truncation at the first order yields the Gaussian approximation.
The second- and third-order corrections are analyzed for the case of small
fugacity and are shown that they substantially improve the
Gaussian-approximation phase diagram. In particular, these corrections
introduce a new conducting phase and make the insulator-conductor coexistence
phase end at the conducting phase. The latter result is in agreement with the
prediction made by generalized renormalization-group
calculations.
\end{abstract}
\vspace{24pt}
PACS numbers : 05.70.Fh; 64.60-i; 64.60.Cn; 68.18.Jk .

Two-dimentional neutral classical Coulomb gas (2DNCCG) can be mapped into
a two-dimensional XY model \cite{1}, and provides a prototype model for
two-dimensional systems with which vortices are important thermal excitations
\cite{2}. These systems including superfluid films, superconducting films,
Josephson-junction arrays, two-dimensional melting, surface roughening,
liquid crystals and double-layer quantum Hall systems \cite{2,3} can undergo
Berezinski\v{i}-Kosterlitz-Thouless (BKT) phase transitions \cite{4}. Hence,
the 2DNCCG has wide applications in studying these systems and, thus, a
thorough understanding of its phase structure is of extreme importance. A
number of analytical and numerical investigations were made to construct
the phase diagram last three decades and they have greatly increased our
understanding on the 2DNCCG. However, the picture is still far from complete
and there remain several issues to be settled as described below.

Both the analytical and simulative investigations have revealed some
universal features of the 2DNCCG. The Gas suffers a BKT phase transition when
the fugacity or particle density is very small or low; that is, for a very
samll fugacity, the gas remains as an insulator with a dipole phase when
temperature is low, whereas, when temperature increases, the dipole charge
becomes unbound, and, thus, the dipole phase becomes a conducting phase. This
kind of phase transition is continuous. Moreover, increasing the fugacity from
a small yet finite value, one encounters a first-order phase transition to an
insulator-conductor coexistence phase.

Nevertheless, the concrete phase diagrams for the 2DNCCG constructed by
various methods have not reached a universal agreement. For example, the
Kosterlitz's renormalization group (RG) equations at the lowest \cite{4}
(1974) and higher orders \cite{5} of an expansion in fugacity showed
existence of just the BKT phase transion only, while Minnhagen's
generalized RG equations \cite{6}, Monte Carlo simulations \cite{7,8},
generalized Debye-H\"{u}ckel-Bjerrum theory \cite{9}, and Gaussian
approximation \cite{10} all led to both the BKT and the first-order phase
transions. Monte Carlo simulations gave even much richer phase structures
\cite{7}. For the connection between the BKT and the first-order phase
transition curves, the Gaussian approximation \cite{10} showed that, in the
fugacity-temperature plane, the first-order phase transition starts from the
end point of the BKT phase transition curve \cite{10}, and
Debye-H\"{u}ckel-Bjerrum theory \cite{9} produced the similar tricritical
point in the density-temperature plane, whereas Minnhagen's investigation
\cite{6} and Monte Carlo simulations \cite{7} indicated that the first-order
phase transition curve passes the end point of the BKT phase transition curve
and ends at a higher temperature and in the conducting phase. Another point of
dispute is the starting point of the BKT phase transition. Although majority
of the calculations \cite{4,5,6,9,10} support the BKT phase transition curve
starts from the point with the zero fugacity (or density) and the reduced
temperature 0.25, we note that in the diagrams obtained from Monte Carlo
simulations \cite{7}, the BKT phase transition curve starts from a slightly
lower, but evidently different temperature. Here, this value of temperature,
0.25 corresponds to Coleman's critical point for the sine-Gordon (sG)
continuum field theory \cite{11}.

The present disagreement on the phase diagram of the 2DNCCG is understandable.
Actually, none of the above methods are exact. The renormalization group
equations \cite{4} (1974) \cite{5} are valid just for low fugacity, so are
Monte Carlo simulations as reviewed in Ref.~\cite{12}. Although an infinite
subset of Feynman diagrams are re-summed up in Minnhagen's procedure \cite{6},
the generalized renormalization-group equations are also valid only for small
fugacities when the temperature is lower than 0.25. Truly, the Gaussian
approximation \cite{10} can work for both small and finite fugacities,
but one does not know how good the results are. As for the generalized
Debye-H\"{u}ckel-Bjerrum theory \cite{9}, it is  of mean-field type
intrinsically.

The above analysis motivates us to propose a systematic, non-perturbative
\footnote{Here, "non-perturbative" means that the coupling of the interacting
system needn't be weak.}, and, at the same time, convergent expansion
scheme on the 2DNCCG to construct the phase diagram. This paper reports our
effort of such an expansion which we call the non-perturbative sG expansion.
Exploiting the equivalence \cite{2,13} between the 2DNCCG and the sG field
theory \cite{11,14}, we perform an expansion on partition function of the
relevant sG field theory in Euclidean space to calculate the thermodynamic
potential of the 2DNCCG up to the third order. The truncated result at the
first order is equal to that obtained from the Gaussian approximation
\cite{10} and, accordingly, the truncated results at higher orders are
expected to provide systematic corrections to the Gaussian-approximation
result. Thus, this scheme can afford an access to the correct phase diagram of
the 2DNCCG. However, the higher-order corrections may give rise to additional
divergences. Therefore, we treat, in the present paper, just the
small-fugacity case for the second- and third-order corrections. In this case,
taking a finite ultraviolet cutoff in the integration over the momentum gets
rid of all the divergences appeared in the truncated expressions. Analysis of
a characteristic parameter which is inversely proportional to the Debye
screening length allows us to obtain phase diagrams of the 2DNCCG up to the
third order. Although our study is limited to the small-fugacity effects of
higher-order corrections, still, it provides systematic improvements towards
the correct phase diagram.

Based on the equivalence between the 2DNCCG and the sG field theory, the
grand canonical partition function for a 2DNCCG consisted of point charges
$\pm q$ \cite{2} \cite{10}(1997) becomes a functional integral over a real
scalar field in the sG formalism,
\begin{equation}
{\cal Z}={\cal Z}_0^{-1}
         \int {\cal D}[\phi] e^{-\int d^2 r {\cal H}_{sG}(\vec{r})} \;,
\end{equation}
with ${\cal Z}_0=\int {\cal D}[\phi]\exp\{\int d^2 r
{\frac {1}{2}}[\nabla\phi(\vec{r})]^2\}.$ The effective Hamiltonian density
for the 2DNCCG, is given by
\begin{equation}
{\cal H}_{sG}(\vec{r})={\frac {1}{2}}[\nabla\phi(\vec{r})]^2
                       -{\frac {2 z}{a^2}}\cos[\beta \phi(\vec{r})] ,
\end{equation}
where $z$ is the electron-self-energy-renormalized fugacity, $a$ an arbitrary
length scale, and $\beta=\sqrt{{\frac {2\pi}{T}}}$ with $T$ the demensionless
reduced temperature. Hereafter, we use $\phi$ instead of $\phi(\vec{r})$.
Calculating the partition function Eq.(1), one obtains the thermodynamic
potential density $\Omega=-T\ln[{\cal Z}]/\int d^2 r$, which is sometimes
called free energy.

Non-perturbative sG expansion, we intend to perform, starts with a
modified Hamiltonian density,
\begin{equation}
{\cal H}_{sG}(\vec{r},\delta)={\cal H}_{\mu} (\vec{r})+\delta \;\;
              {\cal H}_D(\vec{r}),
\end{equation}
which is similar to what Refs.~\cite{15,16,17} did for the $\lambda\phi^4$
field theory, where $\delta$ is a parameter for the convenience of performing
the expansion. Here, ${\cal H}_\mu (\vec{r})=
{\frac {1}{2}}\phi (-\nabla^2+\mu^2) \phi$ and ${\cal H}_D (\vec{r})=
-{\frac {1}{2}}\mu^2\phi^2 - {\frac {2 z}{a^2}} \cos[\beta \phi]$. When
$\delta=1$, the mass parameter $\mu$ cancels out, and, consequently,
${\cal H}_{sG}(\vec{r},\delta)$ is reduced to Eq.(2). Letting
${\cal H}_{sG}(\vec{r},\delta)$ take the place of ${\cal H}_{sG}(\vec{r})$ in
Eq.(1), one can have the following modified partition function
${\cal Z}_\delta$,
\begin{equation}
{\cal Z}_\delta={\cal Z}_0^{-1} {\cal Z}_\mu
         <e^{-\delta\int d^2 r {\cal H}_D(\vec{r})}>_\mu  \;,
\end{equation}
with
\begin{equation}
{\cal Z}_\mu=\int {\cal D}[\phi]\exp\{\int d^2 r {\frac {1}{2}} \phi
    (-\nabla^2+\mu^2) \phi\}
\end{equation}
and
\begin{equation}
<O>_\mu={\cal Z}_\mu^{-1}
         \int {\cal D}[\phi] O \exp\{\int d^2 r {\frac {1}{2}} \phi
    (-\nabla^2+\mu^2) \phi\}
\end{equation}
for an operator $O$. Expanding the exponential function in Eq.(4) into a
Taylor series and, then, making a further expansion according to the series
expression of the function $\ln(1+x)$, one has
\begin{eqnarray}
\ln[{\cal Z}_\delta]&=& -\ln[{\cal Z}_0]+\ln[{\cal Z}_\mu]+
                       \sum_{l=1}^\infty {\frac {(-1)^{(l+1)}}{l}}
                       [\sum_{n=1}^\infty {\frac {(-1)^n}{n!}}
                       \delta^n <(\int d^2 r {\cal H}_D(\vec{r}))^n>_\mu]^l
                       \nonumber  \\
                    &=& -\ln[{\cal Z}_0]+\ln[{\cal Z}_\mu]
                       -\delta <\int d^2 r {\cal H}_D(\vec{r})>_\mu
                       \nonumber  \\
                    && +\delta^2 {\frac {1}{2}}
                        [<(\int d^2 r {\cal H}_D(\vec{r}))^2>_\mu
                        -(<\int d^2 r {\cal H}_D(\vec{r})>_\mu)^2]
                        \nonumber  \\
                    && -\delta^3 [{\frac {1}{3!}}
                          <(\int d^2 r {\cal H}_D(\vec{r}))^3>_\mu
                   +{\frac {1}{3}} (<\int d^2 r {\cal H}_D(\vec{r})>_\mu)^3
                        \nonumber  \\    &&
                        -{\frac {1}{2}}
                        <\int d^2 r {\cal H}_D(\vec{r})>_\mu
                    <(\int d^2 r {\cal H}_D(\vec{r}))^2>_\mu]
                    +\cdots  \;.
\end{eqnarray}
This is a power series in $\delta$. Extrapolating the series to the case of
$\delta=1$ we can get the true thermodynamical potential density $\Omega$
which is independent of $\mu$. However, it is impossible to calculate the
above series exactly, and, therefore, one has to truncate the series at some
order of $\delta$ to obtain an approximation to $\Omega$. In the
approximation, arbitrary $\mu$ does not cancel out any more. Nevertheless,
it can be fixed according to the ``principle of minimal sensitivity''
\cite{18}, which means that, for a credible approximation, $\Omega$ should be
insensitive to variations in $\mu$. The general principle is to minimize the
approximate result with respect to $\mu$. Thus, we have a systematic tool to
approximate $\Omega$ order by order. Note that the above expansion is
similar to the sG expansion in Ref.~\cite{2,6} ($\lambda$ there corresponds
to $\mu$ here) where $\lambda$ remains same for all orders in the expansion.
However, we emphasize that, a crucial distinct point of our expansion is that
the optimized $\mu$ changes from one order to the next. It is this point that
ensures the present expansion convergent \cite{18}. In fact, our expansion is
just a generalization of the scheme proposed in Ref.~\cite{17}. We call it
non-perturbative sG expansion.

Now, we give a concrete calculation of $\Omega$ up to the third order. In
the right hand of Eq.(7), the first two terms are easily calculated by
employing the result of Gaussian functional integral \cite{19}, and one can
find the results in Ref.~\cite{10}. As for the other terms, one can calculate
them with the help of the external source technique of functional
integration \cite{19}. In the calculation process, it is convenient to adopt
the exponential form of cosine function, and the calculations can be performed
as done in Ref.~\cite{10} (1990) and Ref.~\cite{20}. Note that the averages of
multi-cosine products can also be calculated in the same way in Ref.~\cite{10}
(1990) and Ref.~\cite{20}. Thus, a lengthy calculation leads to the following
thermodynamical potential density $\Omega$,
\begin{eqnarray}
\Omega T^{-1}&=&-\int {\frac {d^2 p}{(2\pi)^2}} \ln[p]
      + {\frac {1}{2}}\int {\frac {d^2 p}{(2\pi)^2}} \ln[p^2+\mu^2]
        \nonumber  \\      &&
      - {\frac {1}{2}}\mu^2 I_{(1)}[\mu^2]
      -{\frac {2 z}{a^2}} \exp\{-{\frac {1}{2}}\beta^2 I_{(1)}[\mu^2]\}
         \nonumber   \\ &&
      - {\frac {1}{4}}\mu^4 I^{(2)}[\mu^2]
      +{\frac {1}{2}}{\frac {2 z}{a^2}} \beta^2\mu^2 I^{(2)}[\mu^2]
       \exp\{-{\frac {1}{2}}\beta^2 I_{(1)}[\mu^2]\}  \nonumber  \\  &&
      -{\frac {1}{2}} ({\frac {2 z}{a^2}})^2
       \exp\{-\beta^2 I_{(1)}[\mu^2]\}
       \sum_{k=1}^\infty {\frac {1}{(2k)!}}\beta^{4k}I^{(2k)}[\mu^2]
       \nonumber  \\ &&
       -{\frac {1}{6}}\mu^6 I_{(3)}[\mu^2] +
{\frac {1}{2}}{\frac {2 z}{a^2}}\exp\{-{\frac {1}{2}}\beta^2 I_{(1)}[\mu^2]\}
 [\beta^2\mu^4 I_{(3)}[\mu^2]-{\frac {1}{4}}\beta^4\mu^4 (I^{(2)}[\mu^2])^2]
            \nonumber   \\  &&
     +{\frac {1}{2}} ({\frac {2 z}{a^2}})^2
       \exp\{-\beta^2 I_{(1)}[\mu^2]\}[
      \beta^2\mu^2 I^{(2)}[\mu^2]\sum_{k=1}^\infty {\frac {1}{(2k)!}}
      \beta^{4k}I^{(2k)}[\mu^2]  \nonumber \\  & &
      -\beta^2\mu^2\sum_{k=0}^{\infty}
      {\frac {1}{(2k+1)!}}\beta^{2(2k+1)}] I^{(2+2k+1)}[\mu^2]]
      \nonumber  \\  &&
     -{\frac {1}{2}} ({\frac {2 z}{a^2}})^3
       \exp\{-{\frac {3}{2}}\beta^2 I_{(1)}[\mu^2]\}[
       {\frac {1}{4}}\sum_{i,j=1}^\infty
       {\frac {3 (-1)^{j+1}+1}{2 (2i-1)!j!}}\beta^{2(2i+j-1)}
       I^{(2i-1,j)}[\mu^2] \nonumber  \\
       && + {\frac {1}{4}}\sum_{i,j,k=1}^\infty
       {\frac {(-1)^j}{(2i-1)!j!(2k)!}}\beta^{2(2i+j+2k-1)}
       I^{(2i-1,j,2k)}[\mu^2] \nonumber  \\
       && + \sum_{i,j=1}^\infty
       {\frac {(-1)^{i+j}}{i!j!}}\beta^{2(i+j)}
       I^{(i,j)}[\mu^2]
       +{\frac {1}{3}}\sum_{i,j,k=1}^\infty
       {\frac {(-1)^{i+j+k}}{i!j!k!}}\beta^{2(i+j+k)}
       I^{(i,j,k)}[\mu^2] ]  \;,
\end{eqnarray}
where,
\begin{eqnarray*}
&&I_{(n)}[\mu^2]=\int {\frac {d^2 p}{(2\pi)^2}}[p^2+\mu^2]^{-n}  \;,\\
&&I^{(n)}[\mu^2]=\int \prod_{j=1}^{n-1} {\frac {d^2 p_j}{(2\pi)^2}}
         [p_j^2+\mu^2]^{-1} [(\sum_{j=1}^{n-1} p_j^2)^2+\mu^2]^{-1} \;,\\
&&I^{(2+n)}[\mu^2]=\int {\frac {d^2 p_1}{(2\pi)^2}}[p_1^2+\mu^2]^{-2}
                  [(\sum_{j=1}^{n} p_j^2)^2+\mu^2]^{-1}
                \prod_{j=2}^{n} {\frac {d^2 p_j}{(2\pi)^2}}
             [p_j^2+\mu^2]^{-1}   \;,\\
&&I^{(l,m)}[\mu^2]=(\int d^2 r)^{-1}\int d^2 r_1 d^2 r_2 d^2 r_3
     (\int {\frac {d^2 p_1}{(2\pi)^2}}
     {\frac {e^{i \vec{p_1}\cdot (\vec{r}_1-\vec{r}_2)}}{p^2+\mu^2}})^l
     (\int {\frac {d^2 p_2}{(2\pi)^2}}
     {\frac {e^{i \vec{p_2}\cdot (\vec{r}_2-\vec{r}_3)}}{p^2+\mu^2}})^m
\end{eqnarray*}
and
\begin{eqnarray*}
I^{(l,m,n)}[\mu^2]&=&(\int d^2 r)^{-1}\int d^2 r_1 d^2 r_2 d^2 r_3
     (\int {\frac {d^2 p_1}{(2\pi)^2}}
     {\frac {e^{i \vec{p_1}\cdot (\vec{r}_1-\vec{r}_2)}}{p^2+\mu^2}})^l
     \nonumber  \\  &&
     (\int {\frac {d^2 p_2}{(2\pi)^2}}
     {\frac {e^{i \vec{p_2}\cdot (\vec{r}_2-\vec{r}_3)}}{p^2+\mu^2}})^m
     (\int {\frac {d^2 p_2}{(2\pi)^2}}
  {\frac {e^{i \vec{p_2}\cdot (\vec{r}_2-\vec{r}_3)}}{p^2+\mu^2}})^n \;.
\end{eqnarray*}
In the right hand side of Eq.(8), the first four terms represent the
first-order result, next three terms the second-order correction and all
other the third-order correction.

Eq.(8) has many integrals and series summations. Obviously, many integrals in
Eq.(8) are divergent. Introducing an ultraviolet cutoff in those integrals
can remove the divergences. This is similar to what occurs in the Gaussian
approximation \cite{10,21}. However, those series in Eq.(8) may still not be
convergent for larger $\beta$. In fact, this peculiarity is similar to the
celebrated divergent problem of the sG field theory in the Minkowski
space-time. For the (1+1)-dimensional sG field theory, Coleman's
normal-ordering procedure \cite{11} can get rid of the divergences in the
integrals in perturbation theory \cite{22} and Gaussian approximation
\cite{11,23}. But, at any higher order of the perturbation theory for the
range, $4\pi\le\beta^2\le 8\pi$, some series are still divergent. In the late
1970s and 1980s, several papers proposed a special treatment to the divergent
problem of the sG field theory for the range $4\pi\le\beta^2\le 8\pi$ and
performed further renormalization procedure to turn it finite \cite{22,24},
in addition to Coleman's normal-ordering procedure. As afore mentioned, in
this paper, we do not intend to treat the low-temperature divergent problem in
Eq.(8). We note that in the Gaussian approximation, only linear terms in $z$
survive, and the connection between the transition curves of the first-order
and the BKT phase transitions involves only very small values of $z$. On the
other hand, in Eq.(8), there exist terms with $z^2$ or $z^3$, and only such
terms with $z^2$ or $z^3$ adhere to series summations. Here, we retain only
the linear terms in $z$, and neglect the higher-order terms of $z$ in Eq.(8).
This approximation allows us to compare the present result with the Gaussian
approximation and observe the improvement directly.

Retaining just the linear terms of $z$ in Eq.(8) and carrying out the
integrals with ${\frac {1}{a}}$ as the ultraviolet cutoff, one can get the
dimensionless reduced thermodynamical potential density
$\bar{\Omega}\equiv\Omega a^2 T^{-1}$,
\begin{eqnarray}
\bar{\Omega}&=&{\frac {1}{8 \pi}}\ln[1+{\bar{\mu}}^2]
             -2z(1+{\bar{\mu}}^{-2})^{-{\frac {\beta^2}{8 \pi}}} \nonumber \\
            && -{\frac {1}{16\pi}}{\frac {{\bar{\mu}}^2}{1+{\bar{\mu}}^{2}}}
               +{\frac {z \beta^2}{4\pi}}{\frac {1}{1+{\bar{\mu}}^{2}}}
                (1+{\bar{\mu}}^{-2})^{-{\frac {\beta^2}{8 \pi}}} \nonumber \\
         && -{\frac {1}{48\pi}}{\frac {{\bar{\mu}}^2 (1+2{\bar{\mu}}^{2})}
                {(1+{\bar{\mu}}^{2})^2}}
               +{\frac {z \beta^2}{4\pi}}{\frac {1}{(1+{\bar{\mu}}^{2})^2}}
               (1+{\bar{\mu}}^{-2})^{-{\frac {\beta^2}{8 \pi}}}
       [{\frac {1}{2}} (1+2 {\bar{\mu}}^2)-{\frac {\beta^2}{16 \pi}}]
\end{eqnarray}
with $\bar{\mu}=\mu a$. In Eq.(9), the first two terms are just the Gaussian
approximate result \cite{10}(PRE), the next two terms represent the
second-order correction and the remaining terms the third-order correction.
According to the ``principle of minimal sensitivity'', $\bar{\mu}$ can be
determined by minimizing $\bar{\Omega}$ over $\bar{\mu}$. That is to say,
$\bar{\mu}$ should be satisfied with both the stationary condition (${\frac
{\partial \bar{\Omega}} {\partial {\bar{\mu}}^2}}=0$) and the stabilized
condition (${\frac {\partial^2 \bar{\Omega}}{(\partial {\bar{\mu}}^2)^2}}\ge
0$). Note that $\mu$ is inversely proportional to the Debye screening length
\cite{6,10}, and, accordingly, $\bar{\mu}=0$ implies that the gas becomes
charge-binding and insulating, whereas $\bar{\mu}\not=0$ signifies that the
gas is conducting. Thus, we can obtain the phase structure of the 2DNCCG from
calculated values of $\bar{\mu}$ and $\bar{\Omega}$. We have carried out a
careful analysis of the results order by order. Fig.1 is the $T$$-$$z$ phase
diagrams of the 2DNCCG of various orders constructed from the above
calculation.

In Fig.1, the dotted, dashed and solid curves correspond to the first-, the
second- and the third-order results, repectively (As for the long-dashed line,
see the explanation in the next paragraph). In the first order which is the Gaussian
approximation, the $T$$-$$z$ plane is divided into four regions : I, II, III
and IV. The dotted curve between regions I and II is plotted from the
stationary condition and the critical stabilized condition
${\frac {\partial^2 \bar{\Omega}} {(\partial \mu^2)^2}}= 0$. The dotted curve
between II and III is obtained from $\bar{\Omega}|_{\bar{\mu}=0}=
\bar{\Omega}|_{\bar{\mu}\not=0}$ and coincides with the solid curve in
Fig.1 of Ref.~\cite{10} (PRE). The other two dotted curves which form the
boundaries of region IV are obtained through calculation of $\bar{\Omega}$
point by point and, hence, are rough, but qualitatively correct. In region I,
$\bar{\mu}$ vanishes and the gas consists of dipole pairs. In region II
and III, every point have two values of $\bar{\mu}$, $\bar{\mu}=0$ and $\bar
{\mu}\not=0$, and, accordingly, the gas is in the insulator-conductor
coexistence phase. But in II, the reduced thermodynamical potential
$\bar{\Omega}$ at $\bar{\mu}=0$ is lower than $\bar{\Omega}$ at $\bar{\mu}
\not=0$, and, in III, $\bar{\Omega}$ at $\bar{\mu}=0$ is higher than
$\bar{\Omega}$ at $\bar{\mu}\not=0$. In region IV, $\bar{\mu}\not=0$ and the
gas is in the conducting phase. Analyzing the behaviour of the reduced
thermodynamical potential $\bar{\Omega}$, one can readily see that the dotted
curve between regions I and IV corresponds to a continuous phase transition,
$i.e.$, the BKT phase transition, and on other dotted curves, first-order
phase transitions occur. All of them meet at the tricritical point $\{$T,z$\}=
\{0.25,1/16 \pi\}$.

An added feature in the second and third order results is the region V, which
is separated from the region III by the long-dashed line \footnote{It is
difficult to determine the left boundary of the region V. Here, the
long-dashed line is given just to qualitatively indicate the existence of the
region V.}. Therefore, up to the second and third orders, the dashed and solid
curves with the long-dashed line respectively divide the $T$$-$$z$ plane into
five regions: I, II, III, IV and V. Regions I, II, III and IV have the same
physical meanings respectively with those for the case of the Gaussian
approximation. As for region V, there exist two different non-vanishing
$\bar{\mu}$, and, accordingly, the gas consists in two different kinds of
plasma phases and is a conducting coexistence phase. Note that the long-dashed
line ends at the dashed curve for the second order and at the solid curve for
the third order.

In addition to the introduction of a new phase and the substantial extension
of the insulator-conductor existence region, both the second- and the
third-order corrections change the tricritical point significantly from that
in the Gaussian-approximate phase diagram. The insulator-conductor
coexistence region passes the end point of the BKT curve and end at the
conducting phase. This peculiarity qualitatively agrees with the results from
the Monte Carlo simulations on a triangular and a square lattice \cite{7} and
also from a generalized renormalization-group method \cite{6}. We also
note that the value of $z$ at the lowest point of the dashed curve between
regions II and III is 0.0238 and the corresponding $z$ for the third-order
case is 0.0255. These values show clearly a tendancy of approaching the value
0.0291, which is obtained from the generalized renormalization-group method
\cite{6}.

Comparing our Fig.1 with Fig.1 in Ref.~\cite{10}(PRE), one can see that the
bottom of the dotted curve between regions I and IV, the BKT curve, in this
paper is shifted to the right of the line $T=0.25$, which is given in
Ref.~\cite{10}(PRE). In fact, to get the the BKT critical line,
Ref.~\cite{10} (PRE) made a further approximation to Eq.(7) there. However,
the present curve is obtained by directly checking the value of $\bar{\mu}$
from the Gaussian-approximation expression, without any further analytical
approximation, and the second- and third-order corrections confirm this shift.
We also note that the BKT critical curves obtained from the RG and
generalized RG equations \cite{4,5,6} start from the point $\{T,z\}=
\{0.25,0\}$. Nevertheless, We believe that the BKT critical curves in Fig.1
do not conflict with the Coleman's phase transition point \footnote{The
Coleman's phase transition means that, for the $1+1$-dimensional sG field
continuum, its vacuum-state energy becomes unbounded from below when the
coupling $\beta^2$ increases to $8\pi$, and hence the theory is ill-defined
at that coupling.} and the RG results. Actually, only when the ultraviolet
cutoff is enforced to infinity, the Coleman's phase transition occurs at
$\beta^2=8\pi$ \cite{11}. The RG and the generalized RG equations \cite{4,5,6}
take no such cutoff, whereas the present paper introduces a finite cutoff.
When a finite cutoff is taken, the Coleman's phase transition does not occur
and the sG continuum theory is well-defined at $\beta^2=8\pi$ and even over
\cite{25}.

This paper has performed a non-perturbative sG expansion on the 2DNCCG whose
truncation at the first order is just the Gaussian approximation. We obtained
the phase diagram of the 2DNCCG with the small-fugacity effect of the second-
and the third-order corrections. Our result supports the prediction made
by Minnhagen $et\ al.$ \cite{6} that the insulator-conductor coexistence phase
ends at the conducting phase. Fig.1 shows that the correction to the Gaussian
approximation is important, and the smallness of the third-order correction
presages fast convergency of the expansion in the present scheme. Therefore,
we believe that our result for small fugacities is reliable within the
framework adopted in the present paper, although, as reviewed in
Ref.~\cite{10}(PRE), there is a source for uncertainty owing to the fact that
the mapping between the 2DNCCG and the 2-dimensional sG field theory is exact
only for the Euclidean sG field continuum without any finite cutoff \cite{2}.

In closing the paper, we intend to point out that it is straightforward to
consider further higher order correstions. Moreover, Minnhagen's
RG method \cite{6} is based on a sG expansion, and therefore it is possible
to generalize Minnhagen's method based on the expansion in the present paper.
Perhaps, more interesting and meaningful, a complete treatment to Eq.(8),
$i.e.$, renormalization of the whole expression in Eq.(8) may yield a full
phase diagram of the 2DNCCG. Finally, we intend to emphasize that the scheme
here can be applied to other systems which is relevant to the sG model, for
example, the backward-scattering model, the massive Luttinger model and
Copper Benzoate \cite{25,26,27}, and more general, even to other similar exponential-type interaction systems,
such as, the massive sine-Gordon model \cite{23}, 2D superconductor \cite{28}
and finite-demensional site-disordered spin system \cite{29}.

\acknowledgments
Lu acknowledges Dr. S. Y. Cho, Prof. M. H. Chung and Mr. M. D. Kim for
useful discussions. This project was supported by the Korea Research
Foundation (99-005-D00011). Lu's work was also supported in part by the
National Natural Science Foundation of China under the grant No. 19875034.

\figure{Fig.1 \ \ \
        Comparison among the phase diagrams of the 2DNCCG obtained in the
        Gaussian approximation, and with low-fugacity effects of second- and
        third-order corrections. The dotted, dashed and solid curves
        correspond to the Gaussian-approximation, second- and third-order
        results, repectively. The dotted curves divide the $T$$-$$z$ plane into
        four regions: I,II,III and IV, and both the dashed and the solid
        curves with the long-dashed line yield an additional region V, besids
        regions I, II, III and IV. Region I corresponds to an insulating
        phase, regions IV and V to the conducting phases, and regions II and III
        to the insulator-conductor coexistence phases. Detailed explanation of
        this figure is given in the text.}

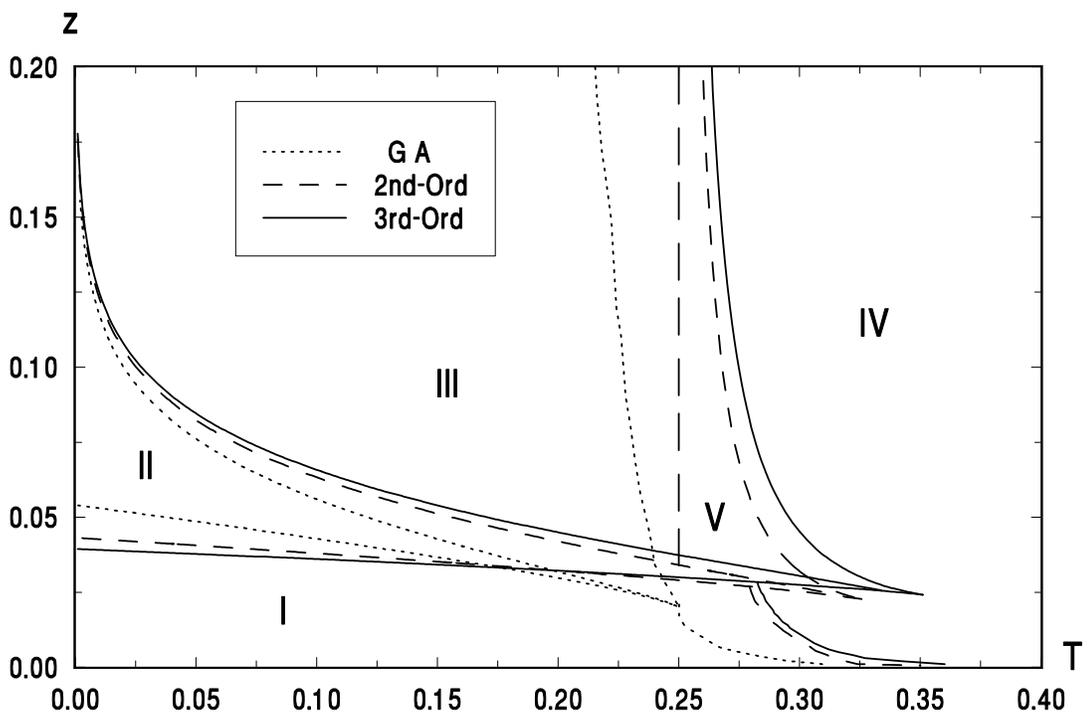
\end{document}